\documentstyle{l-aa}

\def\PRE_CAPSPA{\vspace*{-0.4cm}}
\def\PO_CAPSPA{\vspace*{-0.2cm}}

\input epsf

\begin{document}
\thesaurus{13 (11.02.2; 13.07.2)}

\title{
Measurement of the flux, spectrum, and
variability of TeV $\gamma$-rays from Mkn 501 during a
state of high activity}

\author{
F. Aharonian \inst{1}
\and A.G. Akhperjanian \inst{2}
\and J.A. Barrio \inst{3}\fnmsep\inst{4}
\and K. Bernl\"ohr \inst{1}
\and J.J.G. Beteta \inst{4}
\and S.M. Bradbury \inst{3}\fnmsep\inst{8}
\and J.L. Contreras \inst{4}
\and J. Cortina \inst{4}
\and A. Daum \inst{1}
\and T. Deckers \inst{5}
\and E. Feigl \inst{3}
\and J. Fernandez \inst{3}\fnmsep\inst{4}
\and V. Fonseca \inst{4}
\and A. Fra\ss \inst{1}
\and B. Funk \inst{6}
\and J.C. Gonzalez \inst{4}
\and V. Haustein \inst{7}
\and G. Heinzelmann \inst{7}
\and M. Hemberger \inst{1}
\and G. Hermann \inst{1}
\and M. He\ss \inst{1}
\and A. Heusler \inst{1}
\and W. Hofmann \inst{1}
\and I. Holl \inst{3}
\and D. Horns \inst{7}
\and R. Kankanian \inst{1}\fnmsep\inst{2}
\and O. Kirstein \inst{5}
\and C. K\"ohler \inst{1}
\and A. Konopelko \inst{1}
\and H. Kornmayer \inst{3}
\and D. Kranich  \inst{3}
\and H. Krawczynski \inst{7}
\and H. Lampeitl \inst{1}
\and A. Lindner \inst{7}
\and E. Lorenz \inst{3}
\and N. Magnussen \inst{6}
\and H. Meyer \inst{6}
\and R. Mirzoyan \inst{3}\fnmsep\inst{4}\fnmsep\inst{2}
\and H. M\"oller \inst{6}
\and A. Moralejo \inst{4}
\and L. Padilla \inst{4}
\and M. Panter \inst{1}
\and D. Petry \inst{3}
\and R. Plaga \inst{3}
\and J. Prahl \inst{7}
\and C. Prosch \inst{3}
\and G. P\"uhlhofer \inst{1}
\and G. Rauterberg \inst{5}
\and W. Rhode \inst{6}
\and R. Rivero \inst{7}
\and A. R\"ohring \inst{7}
\and V. Sahakian \inst{2}
\and M. Samorski \inst{5}
\and J.A. Sanchez \inst{4}
\and D. Schmele \inst{7}
\and T. Schmidt \inst{6}
\and W. Stamm \inst{5}
\and M. Ulrich \inst{1}
\and H.J. V\"olk \inst{1}
\and S. Westerhoff \inst{6}
\and B. Wiebel-Sooth \inst{6}
\and C.A. Wiedner \inst{1}
\and M. Willmer \inst{5}
\and H. Wirth \inst{1} \hspace*{0.3cm}(HEGRA Collaboration)
}

\institute{
Max-Planck-Institut f\"ur Kernphysik, P.O. Box 103980,
        D-69029 Heidelberg, Germany
\and Yerevan Physics Institute, Yerevan, Armenia
\and Max-Planck-Institut f\"ur Physik, F\"ohringer Ring 6,
        D-80805 M\"unchen, Germany
\and Facultad de Ciencias Fisicas, Universidad Complutense,
         E-28040 Madrid, Spain
\and Universit\"at Kiel, Inst. f\"ur Kernphysik,
       Olshausenstr.40, D-24118 Kiel, Germany
\and BUGH Wuppertal, Fachbereich Physik, Gau\ss str.20,
        D-42119 Wuppertal, Germany
\and Universit\"at Hamburg, II. Inst. f\"ur Experimentalphysik,
       Luruper Chaussee 149, D-22761 Hamburg, Germany
\and Now at Department of Physics University of Leeds, Leeds LJ2 9JT, UK
}

\offprints{hermann@eu1.mpi-hd.mpg.de}

\date{Received ; accepted}

\maketitle

\markboth{F. Aharonian et al.: Flux, spectrum, and variability
of TeV $\gamma$-rays from Mkn 501}{F. Aharonian et al.: 
Flux, spectrum, and variability
of TeV $\gamma$-rays from Mkn 501}

\begin{abstract}
Between March 16, 1997 and April 14, 1997, a high flux level of 
TeV $\gamma$-rays was observed from Mkn 501, using the HEGRA stereoscopic 
system of four imaging Cherenkov telescopes. The flux level varied during 
this period
from about one half up to six times the flux observed from the Crab Nebula. 
Changes of the detection rate by a factor of up to 4 within 1 day have been 
observed. The measured differential energy spectrum of the radiation 
follows a power law from 1~TeV to 10~TeV.
The differential spectral index
of 2.47 $ \pm 0.07 \pm 0.25$ is close to that of the 
Crab Nebula of $2.66 \pm 0.12 \pm 0.25$.


\end{abstract}

\section{Introduction}

Among the TeV cosmic $\gamma$-ray sources observed by ground-based 
imaging atmospheric Cherenkov telescopes (IACTs) are two 
nearby active
galactic nuclei (AGNs), Mkn 421 
(Punch et al. 1992, Petry et al. 1996) and Mkn 501 
(Quinn et al., 1996, Bradbury et al. 1997). 
In contrast to steady TeV $\gamma$-ray 
sources such as the Crab Nebula, the Whipple group 
(Kerrick et al. 1995a, Gaidos et al. 1996, Buckley et al. 1996) found a 
dramatic time variability in the radiation from Mkn 421, with 
characteristic scales
as short as one hour. 
The study of TeV radiation from these
and other AGNs is interesting for several reasons.
The fast time dependence implies severe limitations on the size of
the source, or on the combination of source size and Doppler
factor in the case of emission from relativistic jets.
TeV $\gamma$-radiation from such distant
objects can furthermore be used to set limits on the diffuse 
extragalactic background radiation at optical and infrared wavelengths
(Biller et al. 1995 and refs given there).
TeV $\gamma$-rays interact with these background radiation 
fields through pair production, with the cross section 
peaking near threshold. The observation of
a cutoff in the $\gamma$-ray spectrum at an energy $E_\gamma$ can be
related to the background photon density at the conjugated
energy $E_{BG} \approx 0.5~\mbox{eV} (1~\mbox{TeV}/E_\gamma)$.

In  early march 1997, the Whipple group communicated the observation of
strong TeV $\gamma$-ray emission from Mkn 501 at a level well above
the flux from the Crab Nebula. Earlier measurements had shown
a flux at a level significantly below the Crab flux. 
Mkn 501 was then detected by the HEGRA IACTs 
CT1 and CT2, operated 
independently from each other, and by CAT (Breslin et al., 1997),
as well as by the
HEGRA IACT system consisting of the four telescopes CT3,4,5,6, 
used in stereoscopic mode. In this Letter, we report the first
results of the analysis of the data obtained with the CT system.
Results obtained with CT1 and CT2 will be reported
elsewhere.

\section{The HEGRA IACT array}

The HEGRA IACT system
(Aharonian 1993) is located on the Canary Island of La Palma, 
at the Observatorio del Roque de los Muchachos
of the Instituto Astrofisico de Canarias,
at a height of about 2200~m asl. 
It consists of six IACTs, the first prototype CT1, the stereoscopic IACT
system (CT3-CT6), and the prototype CT2, which will be included into the 
system after its refurbishment.
The telescopes CT2,4,5,6 are arranged in the corners of a square with 
about 100~m side length, and the telescopes CT3 and CT1 are positioned in
the center of the square. With the stereoscopic system, an air shower
is viewed simultaneously from different directions, allowing to
reconstruct the location of the shower axis in space, and in particular
the direction of the primary and the core location. 
The four telescopes CT3-CT6 are essentially identical; they
were installed during 1995 and 1996 and have been taking data 
as a 4-telescope system since Winter 1996/97. The telescopes have
mirrors with 8.5~m$^2$ area and 5~m focal length and are
equipped with 271-pixel
photomultiplier (PMT) cameras with $0.25^\circ$ pixel size and a $4.3^\circ$
field of view (Hermann 1995). For the trigger of the system a coincidence of
at least two out of four telescopes is required.

\section{Data sample and analysis}

The Mkn 501 data sample comprises data from 14 nights from
March 15/16 to April 13/14, 1997 with a total observation time of
26.7 hours.  Bad weather conditions and the rising moon prevented
continuous observation.
All observations were carried
out in a mode where Mkn 501 was displaced in declination by 
$\pm 0.5^\circ$ from the optical axis of the telescopes, with 
the sign of the displacement changing every 20 min. A region
displaced symmetrically by the same amount in the opposite direction
was used to provide a control sample. 

The image analysis and the reconstruction of the shower axis from
the images is described elsewhere (Daum et al., 1997). In the
present analysis, improved corrections for the telescope
pointing were applied, and an algorithm to estimate the shower
energy was added.
Monte-Carlo simulations were used to determine the
relation between the light yield measured in a camera
as the sum of pixel amplitudes, $Q = Q(r,E)$,
the energy $E$ of the shower, and the distance $r$ to the
shower core. In addition, the fluctuation of the light yield,
$\Delta Q(r,E)$, was determined, taking into account the
error in the measurement of $r$. The shower energy is then
obtained as a weighted average over telescopes. 

The system is expected to provide
a $\gamma$-ray energy threshold of 500~GeV, an energy resolution
of 20\%,
an angular resolution of about $0.1^\circ$, and a determination 
of the shower impact point of about 15~m in each coordinate.
The angular resolution was verified by observations of $\gamma$-rays
from the Crab Nebula (Daum et al. 1997).

\begin{figure}[tb]
\begin{center}
\mbox{
\epsfysize6cm
\epsffile{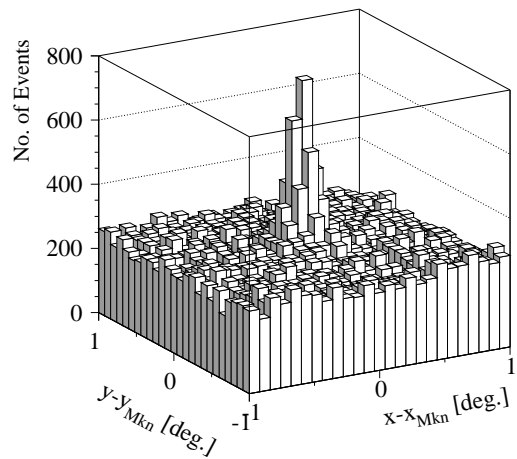}
}
\end{center}
\PRE_CAPSPA
\caption
{Distribution of the reconstructed shower
directions relative to the direction to
Mkn 501, for events where at least
two telescopes triggered, before shape cuts.
\PO_CAPSPA }
\label{fig_2d}
\end{figure}

Already in the raw data, before selection cuts, a clear
signal of Mkn 501 is visible. Fig.~\ref{fig_2d} shows the distribution of
the reconstructed shower directions for all events which triggered
at least two telescopes, and provided two images with 40 or more
photoelectrons and at least two pixels with more than 10 photoelectrons.
The position of Mkn 501, as reconstructed from such distributions
(after cuts on the image shape, to reduce background), is 
consistent with its nominal position within the statistical
error of $0.009^\circ$.

\begin{figure}[tb]
\begin{center}
\mbox{
%\epsfysize7cm
\epsfysize7cm
\epsffile{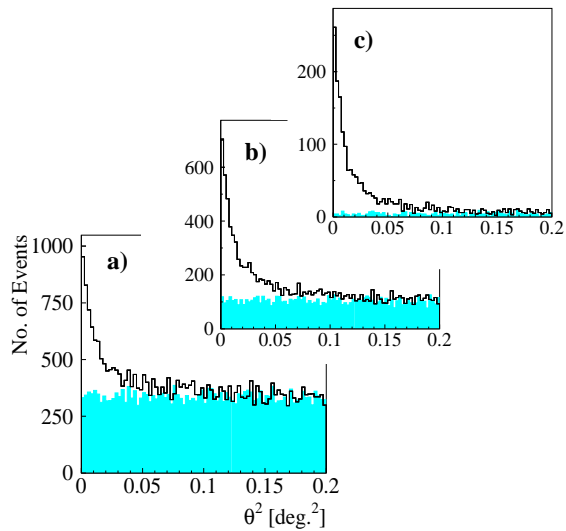}
}
\end{center}
\PRE_CAPSPA
\caption
{Line: distribution $dN/d\theta^2$ of events in the
square of the angle $\theta$ relative to the direction  to
the source. The  shaded histogram shows
the background, see text for details. (a) before cuts,
(b) after loose shape cuts, and (c) after tight shape cuts.
\PO_CAPSPA
}
\label{fig_theta}
\end{figure}

For a quantitative
analysis, 
we plot the distribution in the angle $\theta$
between
the shower axis and the source location; shown in 
Fig.~\ref{fig_theta}(a) is $dN/d\theta^2$. For the uniform
background from charged cosmic rays one expects a flat distribution
in $\theta^2$.
A $\gamma$-ray point source causes an excess around
$\theta \approx 0^\circ$. The observed
distribution shows these features.
An estimate for
the background under the signal is obtained by
plotting the distribution of shower axis relative to
a virtual source displaced by the same amount from the
telescope axis as the real source, but in the opposite
direction. This backgound is shown as a shaded histogram;
it is flat in $\theta^2$. 
In the region up
to $\theta^2 = 0.05^{\circ^2}$ around the source, 3574 excess 
events are counted, corresponding to an average rate of 134 events/h.
The shapes of Cherenkov images can be used to suppress
cosmic-ray background relative to $\gamma$-ray showers;
$\gamma$-rays generate narrower and more compact images.
Therefore the width of each image in a given
event is scaled to the Monte-Carlo expected width of 
$\gamma$-ray images as a function of image amplitude and
distance to the shower core. As selection parameter the mean scaled width
is calculated for all telescopes participating in an event.
To maintain high efficiency and to minimize corrections,
a very loose cut is applied by selecting events with a
mean scaled width below 1.3.
Fig.~\ref{fig_theta}(b) shows the angular distribution
of events after this loose cut. The background is reduced by
a factor of about 3, while the number of events in the peak is
nearly unchanged. We verified that the high selection
efficiency is maintained for all shower energies. 
At the expense of signal statistics, the background can be reduced
further. Fig.~\ref{fig_theta}(c) illustrates the effect of tight cuts
(Daum et al. 1997), which almost completely eliminate the background.

\section{Time variability}
\begin{figure}[tb]
\begin{center}
\mbox{
\epsfysize8cm
\epsffile{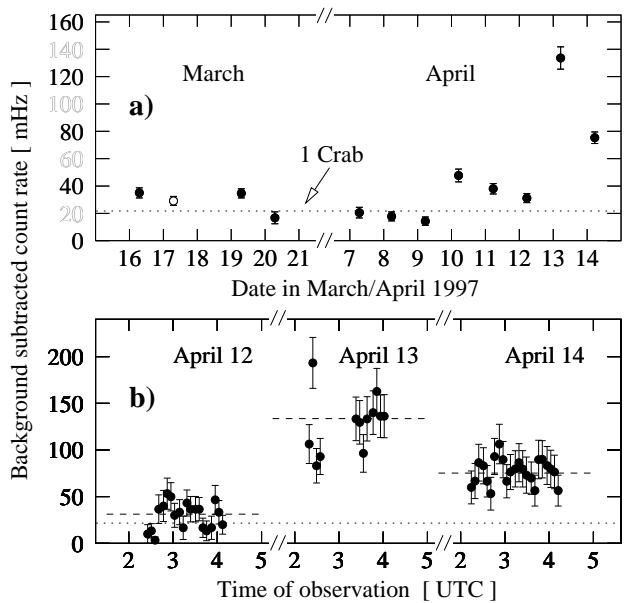}
}
\end{center}
\PRE_CAPSPA
\caption
{Detection rate of Mkn 501 on a night by night basis (a) for
the whole data set and in 5 min. intervals (b) for the last 3 nights.
The dashed lines indicate the average per night, 
the dotted line shows the Crab detection rate. Only observations with a 
zenith angle below $30^\circ$ are considered. On March 17 (open circle)
the trigger rate was reduced by about 15 \% due to slight overcast.
Errors are statistical only. \PO_CAPSPA }
\label{fig_time}
\end{figure}

To investigate time variability, data were grouped in different
time bins, ranging from about 5 min. to entire nights.
We required that at least two telescopes triggered,
and applied a loose angular cut at $\theta = 0.22^\circ$
($\theta^2 = 0.05^{\circ^2})$ as well as the loose selection based on the
widths of the images. Only observations at zenith angles below $30^\circ$ 
with good weather conditions are considered.
Fig.~\ref{fig_time} (a) shows the detection rate on a night by night basis
for the whole data set. While the rate decreases by about 60 \% during
the first 9 nights, it increases by a factor of 3.3 during April 9
and again by a factor of 4.3 during April 12.
Fig.~\ref{fig_time} (b) gives a closer view on the period from
April 12 to April 14 in 5 min. intervals.
Data are statistically consistent with a constant flux within 
each of the 3 nights shown.

\section{Flux and spectral index}

The stereoscopic HEGRA IACT system with its energy resolution of about 
20\% allows detailed studies of the spectrum of 
$\gamma$-ray sources. To derive the flux $F(E)$, the rate at a given
reconstructed energy $E$ is divided by the selection efficiency and the
(energy-dependent) effective detection area. 
Only events where at least
three telescopes triggered are used, to guarantee good
stereoscopic reconstruction. Events are counted 
within a radius of $0.26^\circ$ from the source, 
and the background determined from the average yield
of events around a virtual source on the opposite side of the
camera is subtracted. To avoid significant 
Monte-Carlo based corrections, the loose shape cuts were used, with
a $\gamma$-ray efficiency above 90\%. A similarly conservative approach  
is followed for the effective detection area.  We impose an energy-dependent
limit on the maximum distance $r_{max}(E)$ between the shower core and
the central telescope, such that the trigger probability according
to simulations
is at least 80\%, and also require at least one active telescope
within 140~m from the core. 
Between 1~TeV and 4~TeV, $r_{max}$ rises from about 100~m  to 200~m.
After this selection, the effective area is determined by simple
geometry, up to a small correction.
Below about 0.8~TeV, trigger probabilities do not
safely saturate and data are not used.
In the determination of spectra, only
runs with zenith angles below $30^\circ$ were included,
with a median zenith angle of $18^\circ$.

\begin{figure}[tb]
\begin{center}
\mbox{
\epsfysize7.cm
\epsffile{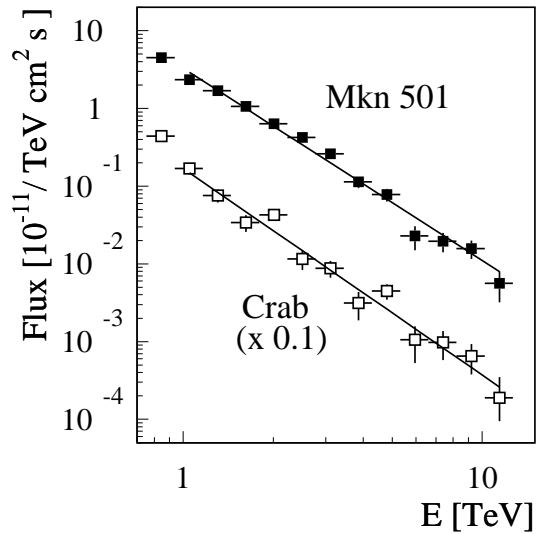}
}
\end{center}
\PRE_CAPSPA
\caption
{Average differential spectrum of $\gamma$-rays from Mkn 501
and from the Crab Nebula. 
The Crab data points are scaled by a factor 0.1. The lines represent 
power-law fits, see text. Only statistical errors are shown.
The energy scale has a 20\% systematic error. \PO_CAPSPA }
\label{fig_flux}
\end{figure}

The resulting differential energy spectrum of Mkn 501 is shown for
energies up to 10 TeV in 
Fig.~\ref{fig_flux},
together with the spectrum of the Crab Nebula analyzed in the
identical fashion, based on 9.7~h of earlier observations at small
zenith angles. The width of
the energy bins corresponds roughly to the rms energy resolution
of about 20\%.
Both spectra are compatible with pure power laws, with a 
differential spectral index of  $2.66 \pm 0.12$ (stat. error 
only) in case of the Crab Nebula, and $2.47 \pm 0.07$ for
Mkn 501; the integral fluxes above 1 TeV are 
$1.0 \pm 0.1 \cdot 10^{-11}$/cm$^2$s
(stat. error only) and 
$2.2 \pm 0.1 \cdot 10^{-11} $/cm$^2$s
, respectively.
We note that this procedure has also been applied on
a night by night basis and that the shape of the spectrum does not
change within the statistical errors of about 0.2. Also it is found 
that the spectrum of Mkn 501 shows no indication for a cutoff in the 
energy range  from 1 TeV to 10 TeV. 
 
To estimate the systematic errors on the flux and the spectral
slopes, the cuts and reconstruction procedures were varied
over a wide range. E.g., the width cut was omitted entirely, or
the angular cut increased to $0.3^\circ$, or the
maximum core radius was limited to 100~m. Different weights and radial
dependencies were used in the energy determination.
The fit range was varied. From these studies, we
estimate a systematic error of $\pm 25$\% in the flux and $\pm 0.25$
in the spectral
slope. An additional error of 36\% on the flux arises from the 20\%
uncertainty in the absolute energy calibration,
increasing the total systematic error on the flux to 45\%.
It is likely that these errors can be reduced as our 
experience in the analysis of IACT system data increases. 
In the comparison of the characteristics of $\gamma$-ray emission
from the Crab Nebula and from Mkn 501, the systematic errors should
cancel to a large extent.

We note, that within the statistical and systematic errors, the measurements
of the $\gamma$-ray flux from the Crab Nebula are consistent
with earlier HEGRA measurements using the single telescopes 
CT1 and CT2 (Konopelko et al. 1996, Petry et al. 1996, Bradbury et al. 1997).

\section{Conclusions}

Observations of Mkn 501 with the stereoscopic HEGRA IACT system during
14 days in March and April 1997 showed a $\gamma$-ray flux at a level
of about one half to six times the Crab flux. The energy spectrum is 
comparable to that of the Crab Nebula, and extends up to at least 10 TeV. 
The flux level was studied on time scales between as short as 5 min and days. 
On a day to day scale an increase of the flux by a factor of up to 4 could 
be observed. On sub-hour time scales the data are statistically 
consistent with a constant flux.
The observations with all 6 HEGRA telescopes are still continued.
The results will be published elsewhere. 

\section*{Acknowledgements}

We appreciate the prompt information from T.C.~Weekes about the
activity of Mkn 501 as observed by the Whipple telescope.
The support of the German Ministry for Research 
and Technology BMBF and of the Spanish Research Council
CYCIT is gratefully acknowledged. We thank the Instituto
de Astrofisica de Canarias for the use of the site and
for providing excellent working conditions. We gratefully
acknowledge the technical support staff of Heidelberg,
Kiel, Munich, and Yerevan.

\end{document}